\documentclass[aps,prl,twocolumn,groupedaddress,superscriptaddress,showpacs]{revtex4}
\usepackage{natbib}
\usepackage[latin1]{inputenc}
\usepackage[dvips]{graphicx}
\usepackage{amsmath,amsbsy,amsfonts,amssymb}
\usepackage{bm}

\newcommand{\state}[3]{\ensuremath{\,^{#1}{#2}_{#3}}}
\newcommand{\unit}[1]{\ensuremath{\,\mathrm{{#1}}}}
\newcommand{\mgp}{\ensuremath{\rm{Mg}^+}}
\newcommand{\mghp}{\ensuremath{\rm{MgH}^+}}
\newcommand{\mgdp}{\ensuremath{\rm{MgD}^+}}
\newcommand{\hh}{\ensuremath{\rm{H}_2}}
\newcommand{\dd}{\ensuremath{\rm{D}_2}}
\newcommand{\hd}{\ensuremath{\rm{HD}}}

\begin{document}
\title{Probing isotope effects in chemical reactions using single ions}
\author{Peter F. Staanum}
\affiliation{QUANTOP - Danish National Research Foundation Centre
for Quantum Optics} \affiliation{Department of Physics and
Astronomy, University of Aarhus, Denmark}
\author{Klaus Højbjerre}
\affiliation{QUANTOP - Danish National Research Foundation Centre
for Quantum Optics} \affiliation{Department of Physics and
Astronomy, University of Aarhus, Denmark}
\author{Roland Wester}
\affiliation{Physikalisches Institut, Universität Freiburg,
Hermann-Herder-Strasse 3, 79104 Freiburg, Germany}
\author{Michael Drewsen}
\affiliation{QUANTOP - Danish National Research Foundation Centre
for Quantum Optics} \affiliation{Department of Physics and
Astronomy, University of Aarhus, Denmark}
\date{\today}
%\received{\today}
%--------------------------------------------------------------------
\begin{abstract}

Isotope effects in reactions between \mgp\ in the
3p\state{2}{P}{3/2} excited state and molecular hydrogen at thermal
energies are studied through single reaction events. From only
$\sim$250 reactions with HD, the branching ratio between formation
of \mgdp\ and \mghp\ is found to be larger than 5. From additional
65 reactions with \hh\ and \dd\ we find that the overall decay
probability of the intermediate MgH$_2^+$, MgHD$^+$ or MgD$_2^+$
complexes is the same. Our study shows that few \emph{single} ion
reactions can provide \emph{quantitative} information on ion-neutral
reactions. Hence, the method is well-suited for reaction studies
involving rare species, e.g., rare isotopes or short-lived unstable
elements.

\end{abstract}
 \maketitle
%--------------------------------------------------------------------

%%% ----------------------------------------------------------------------
Isotope effects often play an important role for the outcome of
chemical reactions. For instance the chemical composition of
interstellar clouds is strongly influenced by isotope effects in
certain reactions~\cite{Millar_2003}. In laboratory experiments,
isotope effects observed in isotopic analogs of chemical reactions
can provide details about the reaction dynamics. Substitution
reactions of the type F+HD and Cl+HD are among the simplest
reactions where isotope effects can be present. They were already
studied when the first chemical lasers were developed, in order to
understand the population inversion mechanism and to identify the
laser transitions~\cite{chemical_lasers}. These studies together
with studies of the F+\dd\ reaction strongly stimulated the whole
field of reactive scattering~\cite{Hu_2006_JCP}. Especially the
resonance effects observed in the F+\hh\ reaction and isotopic
analogs~\cite{Neumark_1984_PRL} have been subject to numerous
experimental and theoretical studies finally resulting in a much
improved understanding of this benchmark
reaction~\cite{Hu_2006_JCP,Qiu_2006_Science}. In another series of
experiments, reactions between a beam of ground state atomic ions
and \hh, HD and \dd\ have been studied. Strong isotope effects have
been observed and, e.g., in reactions between alkaline earth ions
(X$^+$) and HD, it was found that XD$^+$ formation is preferred for
some alkaline earth ions, while XH$^+$ formation is preferred for
others~\cite{Dalleska_1993_JPhysChem,Georgiadis_1988_JPhysChem}.

Within the last few years, new techniques have emerged for
ion-neutral reaction studies, e.g., of cold reactive
collisions~\cite{Wilitsch_2008}, single ion
reactions~\cite{Drewsen_2004_PRL} as well as reaction dynamics by
applying crossed molecular beam imaging~\cite{Mikosch_2008_Science}.
These techniques open up new possibilities in the field of
ion-neutral reactive scattering.

In this Letter, we investigate isotope effects in reactions of
single ions with the isotopologues of molecular hydrogen, a model
system for ion-neutral reactions. More specifically, we consider
reactions at thermal energies with \mgp\ in the 3p\state{2}{P}{3/2}
excited state (excitation energy of 4.4\unit{eV}). Due to the simple
internal structure of the reaction partners, the studied reactions
represent a simple test case for reaction dynamics involving an
electronically excited atomic collision partner. We make use of an
experimental technique with almost 100\% efficiency in detecting
single reaction events~\cite{Drewsen_2004_PRL}. With a total of only
about 300 reactions, the branching ratio between the reactions
\begin{align}
  \rm{Mg}^+(3p\state{2}{P}{3/2})+\rm{HD}&\rightarrow \rm{MgD}^++\rm H \label{eq:MgH-HD} \\
  &\rightarrow \rm{MgH}^++\rm D \label{eq:MgD-HD}
\end{align}
has been found to be larger than 5. In reactions with \hh, HD and
\dd\ the decay paths leading to either \mghp\ or \mgdp\ formation
from a MgH$_2^+$, MgHD$^+$ or MgD$_2^+$ complex have been found to
be equally likely within statistical uncertainties. Our experiments
demonstrate the prospects for similar single molecular ion studies,
e.g., using state prepared molecular ions~\cite{Vogelius_PRL_PRA},
more complex molecular
ions~\cite{Ostendorf_2006_PRL,Hoejbjerre_2007_Aniline} or of
astrophysically relevant
reactions~\cite{Gerlich_2002,Trippel_2006_OH-}. The high detection
efficiency can furthermore be useful for studies of reactions
involving ions of rare species, e.g., superheavy
elements~\cite{Heavy_ions}.

\begin{figure}[!tbp]
\centering\includegraphics[width=0.6\linewidth]{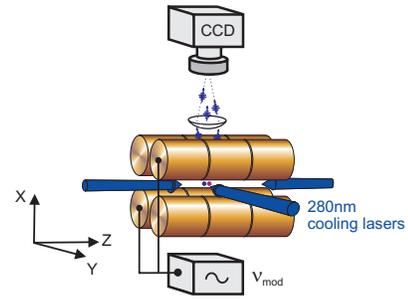}
\caption{Sketch of the experimental setup for non-destructive
identification of molecular ions. See text for
details.}\label{fig:setup}
\end{figure}
In our study we use a linear Paul trap setup which is described in
detail in Ref.~\cite{Drewsen_2003_IJMS}. Briefly, as shown in
Fig.~\ref{fig:setup}, the trap consists of four cylindrical rods,
each sectioned into three parts. By applying suitable ac and dc
voltages (not indicated in Fig.~\ref{fig:setup}) a harmonic
confining potential is created with oscillation frequencies
$\omega_x\approx\omega_y>\omega_z$ along the x-, y- and z-axes,
respectively. $^{26}$\mgp\ ions are loaded into the trap by crossing
an effusive beam of Mg atoms with a laser beam at 285\unit{nm} in
the trap center for resonance-enhanced isotope selective two-photon
ionization of $^{26}$Mg ~\cite{Kjaergaard_2000_ApplPhysB,
Mortensen2004}. The $^{26}$\mgp\ ions are Doppler laser cooled on
the 3s\state{2}{S}{1/2}-3p\state{2}{P}{3/2} transition near
280\unit{nm}. Individual ions are observed by imaging light, emitted
spontaneously during the laser cooling process, onto a
charge-coupled device (CCD) camera.

\begin{figure}[!tbp]
\centering\includegraphics[width=\linewidth]{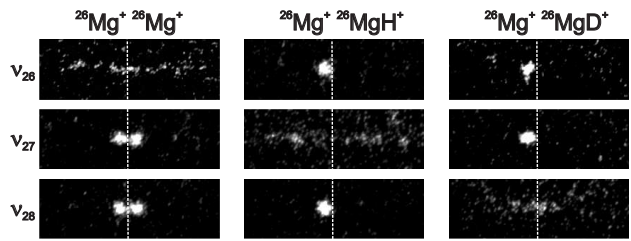}
\caption{CCD images of two-ion Coulomb crystals with modulation
applied at frequencies $\nu_{26}$, $\nu_{27}$ and $\nu_{28}$ for
identification of $^{26}$\mghp\ and $^{26}$\mgdp\ ions. The dashed
vertical lines indicate the position of the trap center along the
z-axis.}\label{fig:IonExcitation}
\end{figure}

Reactions with HD, H$_2$ or D$_2$ molecules are investigated by
first leaking the gasses into the trap chamber until a steady-state
pressure of about $~10^{-9}$\unit{Torr} is reached, and then loading
two $^{26}$\mgp\ ions into the trap. Reactions exclusively take
place with $^{26}$\mgp\ ions excited to the 3p\state{2}{P}{3/2}
state in the laser cooling process, since reactions between
$^{26}$\mgp\ ions in the 3s ground state and thermally excited
hydrogen molecules are not energetically allowed
~\cite{Molhave_2000_PRA}. After a reaction, the formed molecular ion
stays trapped since its acquired kinetic energy only amounts to a
small fraction of the trap depth ($\sim$1 eV). Within tens of
milliseconds after a single molecular ion is formed, it is
sympathetically cooled through the Coulomb interaction with the
remaining laser cooled $^{26}$\mgp\ ion to form a two-ion Coulomb
crystal. Although the molecular ion does not emit light, its
presence is evident from the fact that the remaining laser cooled
$^{26}$\mgp\ ion is located at one of the two initial $^{26}$\mgp\
ion positions (see Fig.~\ref{fig:IonExcitation}).

For unambiguous identification of the molecular ion species, its
mass is determined by applying the identification technique
demonstrated for CaO$^+$ ions in Ref.~\cite{Drewsen_2004_PRL}. In
short, the method relies on exciting the center-of-mass (CM) mode of
the cold two-ion system along the $z$-axis by applying a voltage
oscillating at a frequency $\nu_{mod}$ to two of the end-electrodes
as shown in Fig.~\ref{fig:setup}. When $\nu_{mod}$ is equal to the
eigenfrequency of the CM mode, the motion of the ions is resonantly
excited. This excitation is clearly visible in the CCD-images as a
smearing-out of the fluorescence light from the $^{26}$\mgp\ ion
along the $z$-axis (see Fig.~\ref{fig:IonExcitation}) due to the
long exposure time (100\unit{ms}) compared to the CM mode
oscillation period (typically $\sim$ 10\unit{\mu s}). Since the CM
eigenfrequency depends on the masses of both ions, the mass of the
reaction product ion can now be determined~\cite{Drewsen_2004_PRL}.

In practice, the modulation frequency is scanned in five steps
through a 100\unit{Hz} narrow frequency interval around $\nu_m$,
where $\nu_m$ denotes the CM mode eigenfrequency for one
$^{26}$\mgp\ ion and one singly-charged ion of mass $m$, and at each
step a CCD image is recorded. Such images are shown in
Fig.~\ref{fig:IonExcitation} for one $^{26}\rm{Mg}^+$ ion trapped
simultaneously with another $^{26}\rm{Mg}^+$ ion, a $^{26}$\mghp\
ion and a $^{26}$\mgdp\ ion while modulation frequencies of
$\nu_{26}$, $\nu_{27}$ and $\nu_{28}$ are applied. In the
experiments, modulation frequencies $\nu_{28}$, $\nu_{27}$,
$\nu_{26}$, $\nu_{25}$ and $\nu_{24}$ are applied repeatedly until
one $^{26}$\mgp\ ion has reacted and the reaction product has been
identified. This product ion was $^{26}$\mgdp\ or $^{26}$\mghp\ in
all but two cases where a $^{24}$Mg$^+$ ion produced in a charge
exchange collision was observed (as in Ref.~\cite{Mortensen2004} for
calcium). The choice of working with the less abundant $^{26}$\mgp\
ion instead of the dominant $^{24}\rm{Mg}$ isotope (80\% natural
abundance) was made to avoid problems in distinguishing
$^{24}$MgH$^+$ and $^{24}$MgD$^+$ ions from $^{25}$\mgp\ and
$^{26}$\mgp\ ions, respectively, formed in charge exchange
collisions with background gas Mg atoms \cite{Mortensen2004}. At the
applied molecular gas pressure ($\sim~10^{-9}$\unit{Torr}), the
\mgp-molecule reaction rate is less than one per minute, which means
that the above identification procedure can in general be applied
before both of the initially loaded atomic ions have reacted. After
a reaction product has been identified the trap is emptied, two new
$^{26}$\mgp\ ions are loaded and the experiment repeated.

\begin{figure}[!tbp]
\centering\includegraphics[width=0.9\linewidth]{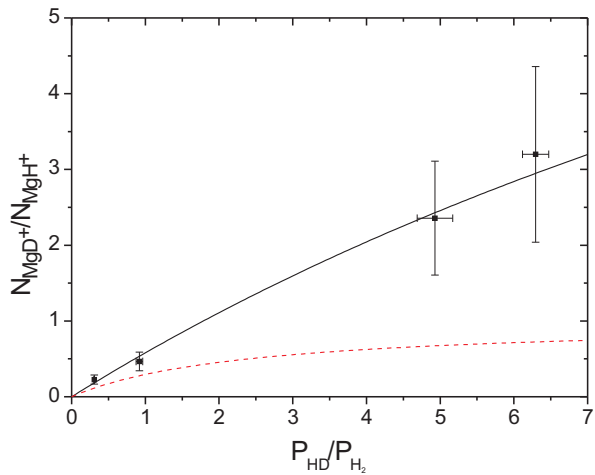}
\caption{The ratio between the number of formed $^{26}\rm{MgD}^+$
and $^{26}\rm{MgH}^+$ ions vs. the relative pressure of HD and H$_2$
gas. From left to right the data points correspond to
$N_{\rm{MgD}^+}=$17, 21, 33, and 32 and $N_{\rm{MgH}^+}=$75, 45, 14,
and 10. The error bars represent statistical uncertainties. A fit of
Eq.~\eqref{eq:ratio} to the data results in
$\eta_{\rm{MgD}^+}^{\rm{HD}}/\eta_{\rm{MgH}^+}^{\rm{H_2}}=0.73\pm
0.17$ and
$\eta_{\rm{MgH}^+}^{\rm{HD}}/\eta_{\rm{MgH}^+}^{\rm{H_2}}=0.06^{+0.10}_{-0.06}$
(black solid line). The red dashed line represents
Eq.~\eqref{eq:ratio} in the absence of isotope effects, i.e. with
$\eta_{\rm{MgD}^+}^{\rm{HD}}/\eta_{\rm{MgH}^+}^{\rm{H_2}}=\eta_{\rm{MgH}^+}^{\rm{HD}}/\eta_{\rm{MgH}^+}^{\rm{H_2}}=0.5$.}\label{fig:H2HD-fit}
\end{figure}

In our study of the branching ratio between reactions
\eqref{eq:MgH-HD} and \eqref{eq:MgD-HD}, the background pressure of
\hh\ is accounted for by measuring the number of $^{26}\rm{MgD}^+$
ions, $N_{\rm{MgD}^+}$, and $^{26}\rm{MgH}^+$ ions,
$N_{\rm{MgH}^+}$, formed at four different ratios of the partial
pressure of HD and H$_2$, $P_{\rm HD}/P_{\hh}$. The obtained results
are presented in Fig.~\ref{fig:H2HD-fit} and show an increase of
\mgdp\ formation with increased HD pressure.

To understand quantitatively the branching ratio between reactions
\eqref{eq:MgH-HD} and \eqref{eq:MgD-HD}, we model the reactions as
two-step processes. In the first step the neutral molecule is
assumed to be captured by the $^{26}$\mgp\ ion at long range to form
a MgH$_2^+$ or MgHD$^+$ collision complex and, in the second step, a
stable \mghp\ or \mgdp\ ion is formed from this complex at short
range. The capture cross-section for the first step is for the
present collision energies well approximated by the Langevin capture
cross section $\sigma=(e/2\epsilon_0)\sqrt{\alpha/(\mu
v^2)}$~\cite{Levine}, where $\alpha$ is the polarizability of the
neutral molecule, $\mu$ is the reduced mass, and $v$ is the relative
velocity in the center-of-mass system. Since the polarizability of
H$_2$ and HD is equal within 1\%~\cite{Kolos_1966_JCP}, in our model
we assume $\alpha$ to be identical for H$_2$ and HD. Due to the low
ion oscillation frequencies in the trap, the reduced masses are not
significantly different from the free-free reaction cases.

The two-step model is valid since the capture range in the first
step is of the order of $(\sigma/\pi)^{1/2}\sim 6$\AA\ which is much
larger than the extension of the hydrogen molecule and the \mgp\ ion
as well as the de Broglie wavelength relevant for the reaction. If
the probability for molecule formation in the second step is denoted
by $\eta$, the molecular ion formation rate $\Gamma$ is then given
by $p_{exc}\times n\times v\times\sigma\times\eta$, where $p_{exc}$
is the $^{26}$\mgp\ excitation probability and $n$ is the neutral
molecule density. Using that density is proportional to pressure we
find that
\begin{align}\label{eq:ratio}
    &\frac{N_{\rm{MgD}^+}}{N_{\rm{MgH}^+}}=\frac{\Gamma_{\rm{MgD}^+}}{\Gamma_{\rm{MgH}^+}}\\\nonumber
    &=\frac{\left(\eta_{\rm{MgD}^+}^{\rm HD}\big{/}\eta_{\rm{MgH}^+}^{\rm H_2}\right)\times\left(P_{\rm {HD}}\big{/}P_{\rm
    H_2}\right)}{\sqrt{\mu_{\rm{HD}}/\mu_{\hh}}+\left(\eta_{\rm{MgH}^+}^{\rm HD}\big{/}\eta_{\rm{MgH}^+}^{\rm H_2}\right)\times\left(P_{\rm {HD}}\big{/}P_{\rm
    H_2}\right)},
\end{align}
where $\eta_{M^+}^{M'}$ is the efficiency of $M^+$ (\mghp\ or \mgdp)
formation after reaction with $M'$ (\hh\ or HD). This expression is
independent of $p_{exc}$ and of the \emph{absolute} partial
pressures, which greatly reduces systematic errors. The relative
partial pressures are measured with a restgas analyzer and for all
values of the ratio $P_{\rm HD}/P_{\hh}$ we estimate its systematic
uncertainty to be less than 25\%. From the parameters determined
from the fit (see Fig.~\ref{fig:H2HD-fit}), the branching ratio
$\eta_{\rm{MgD}^+}^{\rm{HD}}/\eta_{\rm{MgH}^+}^{\rm{HD}}$ can be
determined. Taking the statistical uncertainties into account, we
find that the branching ratio is unbound from above and has a lower
limit of 5, thus demonstrating a dramatic intramolecular isotope
effect. In addition, the ratio
$(\eta_{\rm{MgH}^+}^{\rm{HD}}+\eta_{\rm{MgD}^+}^{\rm{HD}})/\eta_{\rm{MgH}^+}^{\hh}$
is consistent with unity which indicates that the probability of
forming a magnesium hydride ion is equally large from any of the two
complexes MgH$_2^+$ or MgHD$^+$.

In a second series of measurements we studied reactions between
\mgp\ and a mixture of 43$\pm 2$\% \hh\ and 57$\pm 2$\% \dd\ gas. We
observed the production of 40 \mghp\ ions and 25 \mgdp\ ions. Using
the same two-step model as above we find that
$\eta_{\mghp}^{\hh}/\eta_{\mgdp}^{\dd}=1.5\pm0.4$.

Substitution reactions between a $^{26}\rm{MgH}^+$ or
$^{26}\rm{MgD}^+$ ion and a H$_2$, HD or D$_2$ molecule could
potentially give rise to a systematic error. Such reactions were,
however, only observed on two occasions (one
$\rm{MgH}^++\rm{D_2}\rightarrow \rm{MgD}^++\rm{H}+\rm{D}$, one
$\rm{MgD}^++\rm{H_2}\rightarrow \rm{MgH}^++\rm{H}+\rm{D}$) in a
period of more than 20 minutes and hence they do not give rise to
systematic errors on the results presented here.

The most striking of the above results is the strong intramolecular
isotope effect in reactions \eqref{eq:MgH-HD} and \eqref{eq:MgD-HD}.
This finding cannot be explained by a simple statistical model based
on an assumption of an equal probability for populating
energetically accessible states of $^{26}\rm{MgH}^+$ and
$^{26}\rm{MgD}^+$. This assumption only gives rise to
$\eta_{\mgdp}^{\hd}/\eta_{\mghp}^{\hd}\sim 2$ and therefore we
attribute the observed isotope effect to a dynamical mechanism. In
the ion beam experiments of Ref.~\cite{Dalleska_1993_JPhysChem} a
similar isotope effect has been observed in reactions between
\emph{ground state} \mgp\ ions and HD molecules at center-of-mass
energies up to 11\unit{eV}. This was rationalized in terms of an
impulsive interaction with a thermodynamic threshold.

\begin{figure}[!tbp]
\centering\includegraphics[width=\linewidth]{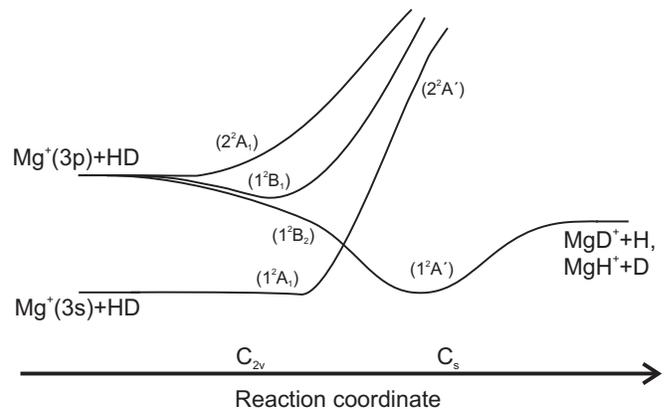}
\caption{Sketch of relevant potential surfaces for the \mgp+\hd\
reaction~\cite{Kleiber_1998_IntRevPhysChem,Bauschlicher_1993_CPL}.
On the left (right) hand side the potential surfaces are labelled
according to C$_{2v}$ (C$_s$) symmetry. As discussed in the text,
the reaction probably proceeds by insertion of \mgp\ into the HD
bond on the 1$^2B_2$ potential surface, followed by \mgp-D or \mgp-H
bond formation.}\label{fig:PotSurfaces}
\end{figure}

A schematic view of the potential surfaces involved in the reaction
is shown in Fig.~\ref{fig:PotSurfaces}. The analogous potential
surfaces for MgD$_2^+$ was explored in a photofragmentation study
where \mgdp\ formation was observed after laser excitation from the
1\state{2}{A}{1} state to the red of the Mg$^+$(3p)+D$_2$
asymptote~\cite{Ding_1993_JPC}. From the observed \mgdp\ spectrum it
was argued that \mgdp\ is formed by direct and fast reactions on the
1\state{2}{B}{2} surface in C$_{2v}$ geometry through a bond-stretch
mechanism as well as from the 1\state{2}{B}{1} state, possibly
through a coupling to the 1\state{2}{B}{2} state. On the
1\state{2}{B}{2} potential surface the \mgp\ ion becomes inserted in
the \dd\ bond such that the D-D bond is stretched and eventually
broken and a \mgp-D bond is
formed~\cite{Kleiber_1998_IntRevPhysChem,Ding_1993_JPC}. In
addition, in this study no dissociation into \mgp(3s)+\dd\ was
observed. In our two-step model this corresponds to the values of
$\eta_{\mghp}^{\hh}$, $\eta_{\mgdp}^{\dd}$ and
($\eta_{\mgdp}^{\hd}+\eta_{\mghp}^{\hd}$) being close to unity,
which indeed is in good agreement with the \mgp+\hh\ reaction rate
measured in a previous
study~\cite{Molhave_2000_PRA,Molhave_speciale}.

The investigation of photofragmentation indicate that the \mgp+HD
reaction discussed in the present paper proceeds via the
1\state{2}{B}{2} surface through a bond-stretch mechanism that
eventually favors the formation of
\mgdp~\cite{Kleiber_1998_IntRevPhysChem,Ding_1993_JPC}. To fully
understand the transition from a MgHD$^+$ complex to a potential
surface favoring the \mgdp +H asymptote rather than the \mghp+D
asymptote requires a detailed theoretical study. It might be
necessary to consider the details of the conical intersection which
arises from the crossing of the 1$^2$A$_1$ and 1$^2$B$_2$ potential
surfaces. Non-adiabatic couplings at the conical intersection could
give rise to a preference of the \mgdp\ channel over the \mghp\
channel. The same mechanism could be responsible for the isotope
effect observed in reactions with ground state \mgp\
ions~\cite{Dalleska_1993_JPhysChem}.

In conclusion, we have found that reactions between \mgp\ in the
3p\state{2}{P}{3/2} excited state and HD molecules at thermal
energies preferentially leads to the formation of \mgdp\ rather than
\mghp\ with a branching ratio larger than 5. Additional reactions
with \hh\ and \dd\ have shown that after a reaction complex has
formed through a Langevin capture process, the molecular ion
formation efficiencies after capture of a \hd, \hh\ or \dd\ molecule
are equal within statistical uncertainties. The efficiencies are
consistent with unity which is in agreement with observations in a
previous study of MgD$_2^+$ photofragmentation~\cite{Ding_1993_JPC}.
Our measurements demonstrate that it is possible to determine
\emph{quantitatively} the branching ratios and relative reaction
rates in ion-neutral reactions by observing only a few hundred
\emph{single} reactions. The method should in the future be
applicable to a variety of studies, e.g., of astrophysically
relevant reactions, chemistry of superheavy elements and reactions
involving more complex molecules.

\begin{acknowledgments}
The authors are grateful to Jeppe Olsen and Lorenz Cederbaum for fruitful discussions.
PFS acknowledges support from the Danish Natural Science Research
Council. RWs visit in Århus was supported by the European Science
Foundation through the CATS network.
\end{acknowledgments}
%%% ----------------------------------------------------------------------

\end{document}